\documentclass{resonance}
\usepackage{graphicx}
\usepackage{epsfig}
\usepackage{dcolumn}
\usepackage{bm}
\usepackage{amsmath,amssymb,color}
\usepackage{setspace}

\begin{document}

\newcommand{\nwc}{\newcommand}
\nwc{\vs}{\vspace}
\nwc{\hs}{\hspace}
\nwc{\la}{\langle}
\nwc{\ra}{\rangle}
\nwc{\lw}{\linewidth}
\nwc{\nn}{\nonumber}
\nwc{\tb}{\textbf}
\nwc{\td}{\tilde}
\nwc{\Tr}{\tb{Tr}}
\nwc{\dg}{\dagger}

\nwc{\pd}[2]{\frac{\partial #1}{\partial #2}}
\nwc{\zprl}[3]{Phys. Rev. Lett. ~{\bf #1},~#2~(#3)}
\nwc{\zpre}[3]{Phys. Rev. E ~{\bf #1},~#2~(#3)}
\nwc{\zpra}[3]{Phys. Rev. A ~{\bf #1},~#2~(#3)}
\nwc{\zjsm}[3]{J. Stat. Mech. ~{\bf #1},~#2~(#3)}
\nwc{\zepjb}[3]{Eur. Phys. J. B ~{\bf #1},~#2~(#3)}
\nwc{\zrmp}[3]{Rev. Mod. Phys. ~{\bf #1},~#2~(#3)}
\nwc{\zepl}[3]{Europhys. Lett. ~{\bf #1},~#2~(#3)}
\nwc{\zjsp}[3]{J. Stat. Phys. ~{\bf #1},~#2~(#3)}
\nwc{\zptps}[3]{Prog. Theor. Phys. Suppl. ~{\bf #1},~#2~(#3)}
\nwc{\zpt}[3]{Physics Today ~{\bf #1},~#2~(#3)}
\nwc{\zap}[3]{Adv. Phys. ~{\bf #1},~#2~(#3)}
\nwc{\zjpcm}[3]{J. Phys. Condens. Matter ~{\bf #1},~#2~(#3)}
\nwc{\zjpa}[3]{J. Phys. A: Math theor  ~{\bf #1},~#2~(#3)}

\newcommand\bea{\begin{eqnarray}}
\newcommand\eea{\end{eqnarray}}
\newcommand\beq{\begin{equation}}  
\newcommand\eeq{\end{equation}}
\newcommand{\new}{\newpage}
\newcommand{\noi}{\noindent}
\newcommand{\bib}{\bibitem}
\newcommand{\cosec}{\operatorname{cosec}}
\newcommand{\non}{\nonumber}  
\newcommand\mb{\mathbf}
\newcommand\bs{\boldsymbol}
\newcommand\mT{\mathcal{T}}
\newcommand\dd{\text{d}}
\newcommand\s{\sigma}
\newcommand{\p}{\tilde{\Psi}}
\newcommand{\ps}{\tilde{\Phi}}
\newcommand{\C}{\tilde{c}_{j}}
\newcommand{\X}{\tilde{\chi}_{{\bs k}}}
\newcommand\ie{{\it{i.e.}}}
\newcommand\etal{{\it{et al.}}}
\newcommand\eg{{\it{e.g.}}}\def\bbraket#1{\mathinner{\langle\hspace{-0.75mm}\langle{#1}\rangle\hspace{-0.75mm}\rangle}}
\def\i{\imath}
\def\v{\upsilon_F} 
\def\nn{\nonumber}
\def\f{\frac}
\def\al{\alpha}
\def\om{\omega}
\def\de{\delta}
\def\ep{\epsilon}
\def\ga{\gamma}
\def\si{\sigma}
\def\Do{\partial}
\def\De{\Delta}
\def\mb{\mathbb}
\def\mc{\mathcal}
\def\vr{\varrho}
\def\d{\cdot}
\def\t{\tilde}
\def\l{\lambda}
\def\la{\langle}
\def\ra{\rangle}
\def\mbb{\mathbb}
\def\Y{\Upsilon}
\def\ua{\uparrow}
\def\da{\downarrow}
\def\sf{\textsf}
\def\al{\alpha}
\def\be{\beta}
\def\til{\tilde}
\def\ka{\kappa}
\def\sX{\small{X}}
\def\br{{\bs r}}
\def\bk{{\bs k}}

\title{Higher Order Topological Systems: A New Paradigm}
\author{Arijit Saha and Arun M. Jayannavar}




\maketitle{}
\vskip -0.4cm
\authorIntro{\includegraphics[width=3.5cm]{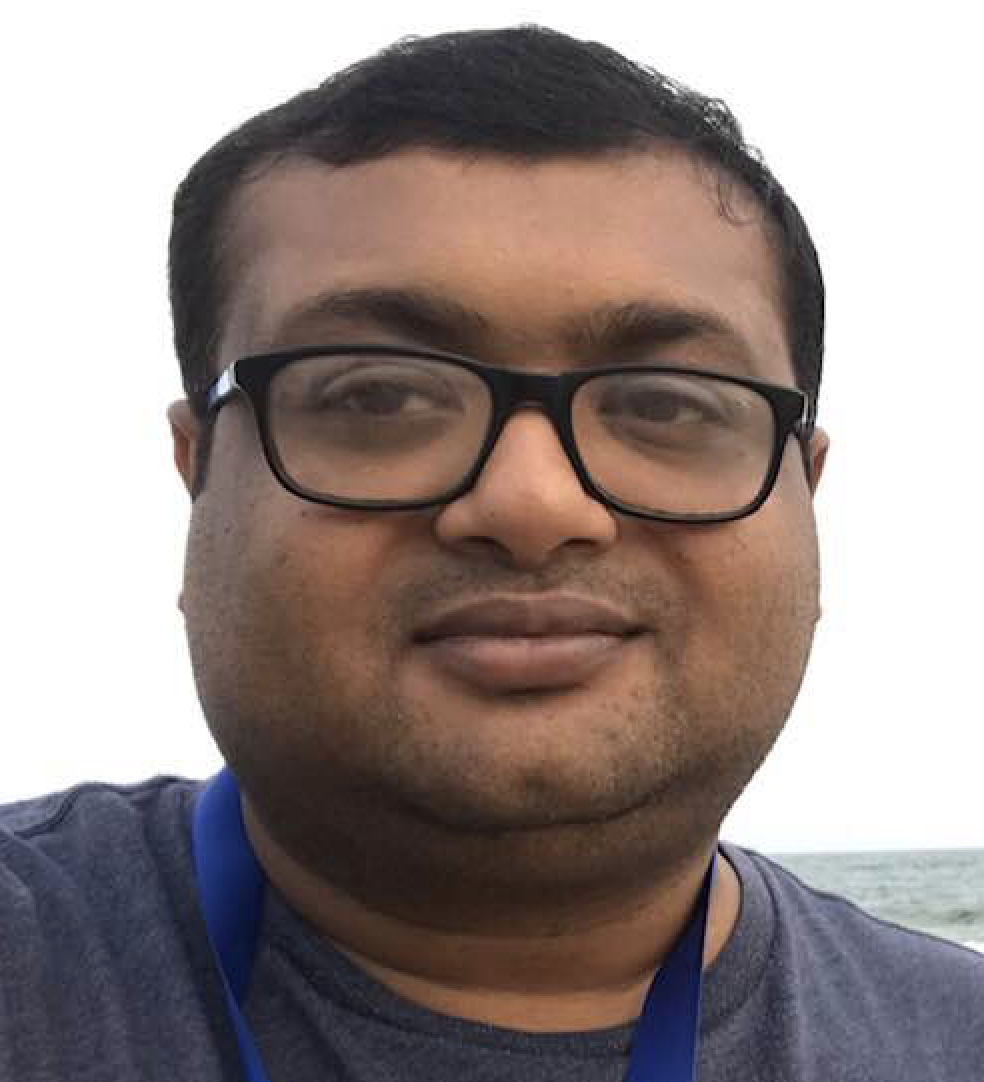}\\
Arijit Saha is a Reader-F at Institute of Physics, Bhubaneswar. His research interest lies broadly in the 
areas of mesoscopic physics, topological insulators, Dirac materials and strongly correlated electrons.
\vskip +0.5cm
\includegraphics[width=3.5cm]{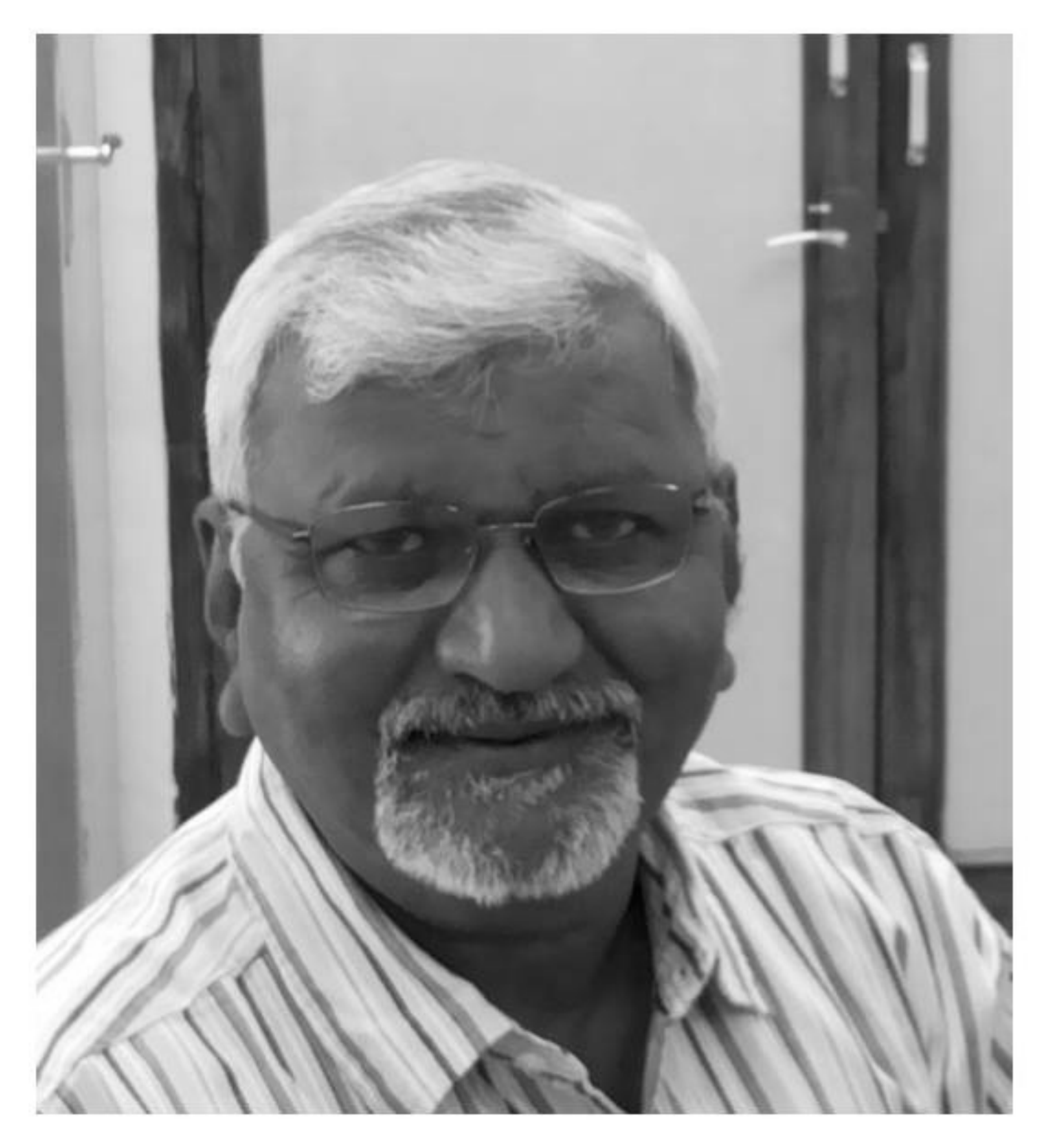}\\
Arun M. Jayannavar is a Senior Professor at Institute of Physics, Bhubaneswar. His research interest lies broadly in 
different aspects of mesoscopic physics and statistical mechanics.}

\begin{abstract}
Higher order topological insulators are a new class of topological insulators in dimensions $\rm d>1$. These higher-order” topological insulators possess $\rm (d - 1)$-dimensional boundaries that, 
unlike those of conventional topological insulators, do not conduct via gapless states but instead are themselves topological insulators. Precisely, an $\rm n^{\rm th}$-order topological insulator in $\rm m$ dimensions hosts $\rm d_{c} = (m - n)$-dimensional boundary modes $\rm (n \leq m)$. For instance, a three-dimensional second (third) order topological insulator hosts gapless modes on the hinges (corners), characterized by $\rm d_{c} = 1 (0)$. Similarly, a second order topological insulator in two dimensions only has gapless corner states ($\rm d_{c} = 0$) localized at the boundary. These higher order phases are protected by various crystalline symmetries. Moreover, in presence of proximity induced superconductivity and appropriate symmetry breaking perturbations, the above mentioned bulk-boundary correspondence can be extended to higher order topological superconductors hosting Majorana hinge or corner modes. Such higher-order systems constitute a distinctive new family of topological phases of matter which has been experimentally observed in acoustic systems, multilayer $\rm WTe_{2}$ and $\rm Bi_{4}Br_{4}$ chains. In this general article, the basic phenomenology of 
higher order topological insulators and higher order topological superconductors are presented along with some of their experimental realization.
\end{abstract}
  
\monthyear{June 2021}
\artNature{GENERAL ARTICLE}

\section{Introduction}
The advent of topological insulators (TIs) has emerged as a new field of research in modern condensed-matter physics both from theoretical and experimental point of view~\cite{qi2010quantum,
moore2009topological, maciejko2011,hasanmoore2010,hasan2010colloquium,sczhangreview,sahajayan}. The realization of topological insulating phases is based on the spin-orbit coupling present in  \keywords{Crystalline topological insulator, topological superconductor, Majorana fermions etc.} a given material~\cite{konig2007quantum,dheish2008,yxia2009,hzhang2009,kanemele2005,bernevig2006quantum,fukane2008}. Precisely, a topological insulator (TI), like an ordinary band insulator, has a bulk energy gap separating the filled valence electronic band from the empty conduction band. However, the two dimensional (2D) surface or one-dimensional (1D) edge of a TI, necessarily has gapless conducting states that are protected by time-reversal symmetry (TRS). This constitutes the topological bulk–boundary correspondence. These gapless boundary modes cannot be removed by local boundary perturbations without breaking the underlying symmetry, thus the system is called TI. In addition to their fundamental interest, these states are predicted to have special properties that could be useful for applications ranging from spintronics to quantum computation~\cite{pesinmacdonald2012,ady1}. Soon after the discovery of TIs, the above mentioned bulk-boundary correspondense has been generalized to 
topological superconductors (TSCs) hosting Majorana zero-modes (MZMs) at their boundaries~\cite{fukane2008,jalicea1,beenakkerreviewMF,stiwary1,lejinseflensberg2012,kitaev}. 

\hskip -7cm
\fbox{\begin{minipage}{16em}
Time reversal symmetry is the discrete symmetry of physical laws under the transformation of time reversal, \ie~$t \rightarrow -t$. In quantum mechanics TRS is represented by an anti-unitary operator
which commutes with the Hamiltonian if the system possesses that symmetry. For instance, in case of spin $\frac{1}{2}$ system, TRS operator can be written as $\hat{T}=-i\eta \sigma_{y}K$
where $\eta$ stands for an arbitrary phase, $\sigma_{y}$ is the Pauli matrix and $K$ is the complex conjugation operator. 
\end{minipage}}
\vskip -6.2cm
Very recently, a new class of topological phases has been introduced to which the usual concept of the bulk-boundary correspondence does not apply~\cite{Schindler,Benalcazar1,Benalcazar2,Parameswaran}.  Here, the topology of the bulk protects gapless states on the hinges (corners), while the surfaces (edges) are gapped. Both systems, 
with gapless corner and hinge states, respectively, can be identified under the notion of higher-order TIs (HOTI)~\cite{Schindler}. To be precise, an $n^{\rm th}$ order TI in $d$-dimension has gapless 
states that dwell on $(d - n)$-dimensional boundaries. For instance, in three dimensions (3D), a second-order TI (SOTI) exhibits gapless states that are located on one-dimensional (1D) ``hinges" 
between distinct gapped surfaces, whereas a third-order TI (TOTI) has gapless states on its zero-dimensional (0D) ``corners''. Similarly, a SOTI in two dimensions (2D) also exhibits gapless 0D corner 
states while the edges remain gapped. The topological character of these HOTI is protected by various bulk crytalline symmetries~\cite{Geier, Luka} for \eg,~product of rotational symmetry and TRS 
($\hat{\mathcal{C}}_{4}$$\hat{T}$)~\cite{Schindler}, inversion symmetry~\cite{Khalaf} etc. Moreover, the intriguing concept of HOTI can be generalized to higher-order topological superconductors (HOTSC) hosting 0D Majorana corner modes (MCMs) and 1D Majorana hinge modes (MHMs) protected by particle-hole (PH) symmetry~\cite{Geier,ZYan,YJWu,RXZhang}. These distinct, recently predicted HOTI phases have been experimentally realized in acoustic systems~\cite{HXue}, $\rm Bi$~\cite{FSchindler}, multilayer $\rm WTe_{2}$~\cite{YBChoi} and Bismuth-halide ($\rm Bi_{4}Br_{4}$) chains~\cite{Noguchi}.

\section{First Order Topological Systems {\label{sec:II}}}
\begin{figure}[!thpb]
\centering
\includegraphics[width=1.0\linewidth]{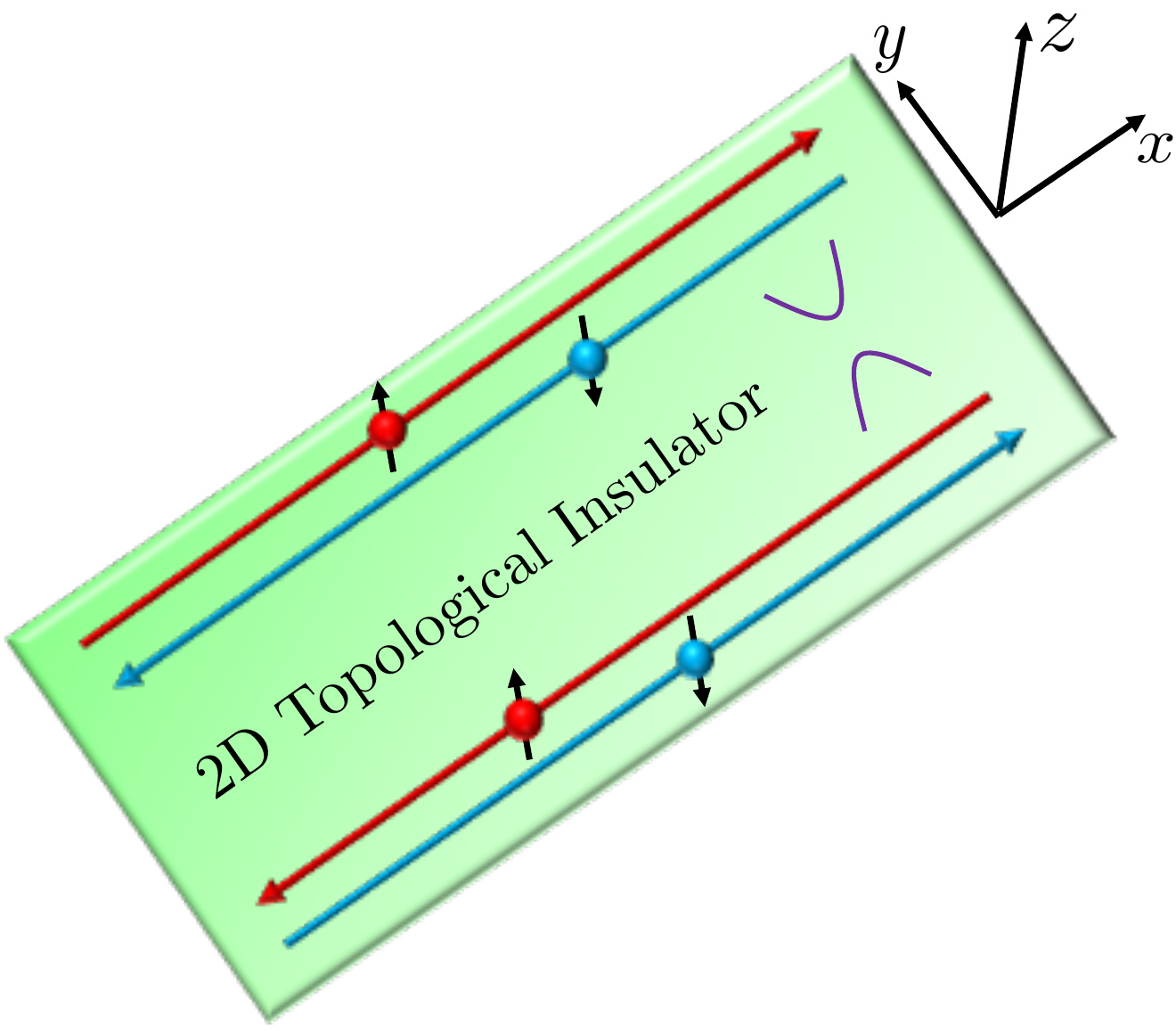}
\vskip-2.8cm
\caption{(Color online)  Cartoon of a 2D TI in which the bulk is gapped and both the 1D egdes are spinful and spin-momentum locked
\ie~of helical nature. The upper edge contains a forward mover with $\uparrow$ spin and a backward mover with $\downarrow$ spin. The spin and momentum 
direction is reversed for the lower edge.
}
\label{Fig1}
\end{figure}

During the last decade, the phenomena of TIs emerge in certain materials with strong spin-orbit coupling (SOC) that preserves TRS. A 2D TI or quantum spin Hall (QSH) state is invariant 
under time reversal, has a charge excitation gap in the bulk, but has topologically protected 1D edge states at the boundary~\cite{qi2010quantum} (see Fig.~\ref{Fig1}). 
\vskip -6cm
\hskip +10.5cm
\fbox{\begin{minipage}{15em}
SOC arises in a material due to broken inversion symmetry as well as crystal symmetry. It is a relativistic effect and acts like an internal magnetic field without violating the TRS. 
\end{minipage}}
\vskip +5cm
Such edge states are called helical as the spin is correlated with the direction of motion. In Fig.~\ref{Fig1}, the upper edge contains a forward (right) mover with $\uparrow$ spin and a backward (left)
mover with $\downarrow$ spin. The spin and momentum directions are reversed for the lower edge. These edge states appear as Kramers doublets due to TRS and are robust against local perturbations
like non-magnetic impurity. The basic mechanism behind the appearance of such edge states is band inversion, in which the usual ordering of conduction band and valence band is inverted by SOC. 
This phenomena was first theoretically predicted in graphene (Kane-Mele model)~\cite{kanemele2005} and then in Mercury-Telluride ($\rm HgTe$) (Bernevig-Huges-Zhang model)~\cite{bernevig2006quantum}. The signature of 2D TI was experimentally observed in HgTe quantum well where in the topological phase, conductance appears to be quantized ($2e^{2}/h$) as the two 
edge states of TI act as two conducting 1D channels contributing $e^{2}/h$ each~\cite{konig2007quantum}. Extending this bulk-boundary correspondence for 3D TI, one obtains spin-momentum locked 
2D surface states at the boundary. These surface states consist of 2D massless Dirac fermions and the corresponding dispersion forms a single massless (zero band gap) Dirac cone. The latter was experimentally observed in bismuth selenide ($\rm Bi_{2}Se_{3}$) using angle resolved photo emission spectroscopy (ARPES) technique~\cite{dheish2008}. 
\vskip -8.5cm
\hskip -6.8 cm
\fbox{\begin{minipage}{15em}
Graphene is a 2D hexagonal lattice made of Carbon atoms. The low energy dispesion relation of graphene can be described by the massless Dirac spectrum $\epsilon_{k}=\hbar v_{F} \lvert k \rvert$.
Here $v_{F}\approx 3\times 10^{6}~\rm m s^{-1}$ is the Fermi velocity and $\lvert k \rvert = \sqrt {k_{x}^{2} + k_{y}^{2}}$. 
\end{minipage}}
\vskip +4.5cm
The study of TI was generalized to topological superconductors (TSC) hosting Majorana fermions (MFs)~\cite{qi2010quantum,jalicea1,beenakkerreviewMF}. The latter is believed to be the basic building
block of future topological quantum computer, which would be exceptionally well protected from errors or decoherence~\cite{ady1}. Fu and Kane first showed the appearence of MF at the vortex core of a TSC which can be engineed in a 3D TI, kept in close proximity to a $s$-wave superconductor and ferromagnetic insulator~\cite{fukane2008}. However, the recent interest in MF has turned into 1D systems~\cite{kitaev} due to their realizibility in semiconducting heterostructures~\cite{jdsau2} and 1D semiconducting nanowires (NW) with strong SOC~\cite{oregetal}. Under suitable circumstances, such NW becomes a TSC and a pair of MZMs appear at the two ends of the NW. In very recent transport measurements~\cite{VMourik,Adas,HZhang}, zero-bias tunneling conductance exhibits a quantized 
conductance plateau at $2e^{2}/h$~\cite{HZhang} when the Majorana mode is present, and no peak when it is absent. Such zero-bias conductance can be interpreted as an indirect experimental evidence for the Majorana zero mode.
\vskip -8.5cm
\hskip -6.8 cm
\fbox{\begin{minipage}{15em}
MFs are zero energy state having the remarkable property of being their own antiparticles. In nanoscience and condensed-matter physics, being its own antiparticle means that a MF must be an equal superposition of an electron and a hole state. For a spinless $p$-wave superconductor, the quasi-particle creation and annihilation operator at zero-energy satisfies the mathematical relation
$\gamma = \gamma^{\dagger}$ \ie~particle being it's own anti-particle. 
\end{minipage}}
\vskip +1.5cm
Note that, so far we discuss about systems which are insulators in their $d$-dimensional interior (bulk) but allow metallic conduction on their $(d - 1)$-dimensional boundaries. Hence, they are known 
as ``First order topological systems''. 

\section{$\rm 2D$ Higher Order Topological Insulator {\label{sec:III}}}
Before we begin discussing about the basic phenomenology of HOTI, here we present a basic picture of zero-energy Jackiw-Rebbi (JR) modes~\cite{JackiwRebbi} that will be necessary to understand 
the emergence of higher-order topological modes in these systems. 
\subsection{A primer on Jackiw-Rebbi zero modes}
Let us consider a massive 1D Dirac Hamiltonian of the form
\begin{eqnarray}
\mathcal{H}=-i\hbar v_{F}\sigma_{z}\partial_{x} - m v_{F}^{2}\sigma_{x}\ ,
\label{Eq1}
\end{eqnarray}
where $\sigma_{j}$ for $j \in \{x, y, z\}$ are Pauli matrices acting on the spin space, $v_{F}$ is the Fermi velocity and $m$ is the mass term. The bulk spectrum of this system is given by
$E_{\pm}(k)=\pm \sqrt{(\hbar v_{F} k)^{2} + m^{2}}$ which is gapped (alike an ordinary insulator) as $m\neq 0$. We now allow the mass term to become spatial-dependent, \ie~$m=m(x)$.
\vskip +0.5cm
\begin{figure}[!thpb]
\centering
\includegraphics[width=1.0\linewidth]{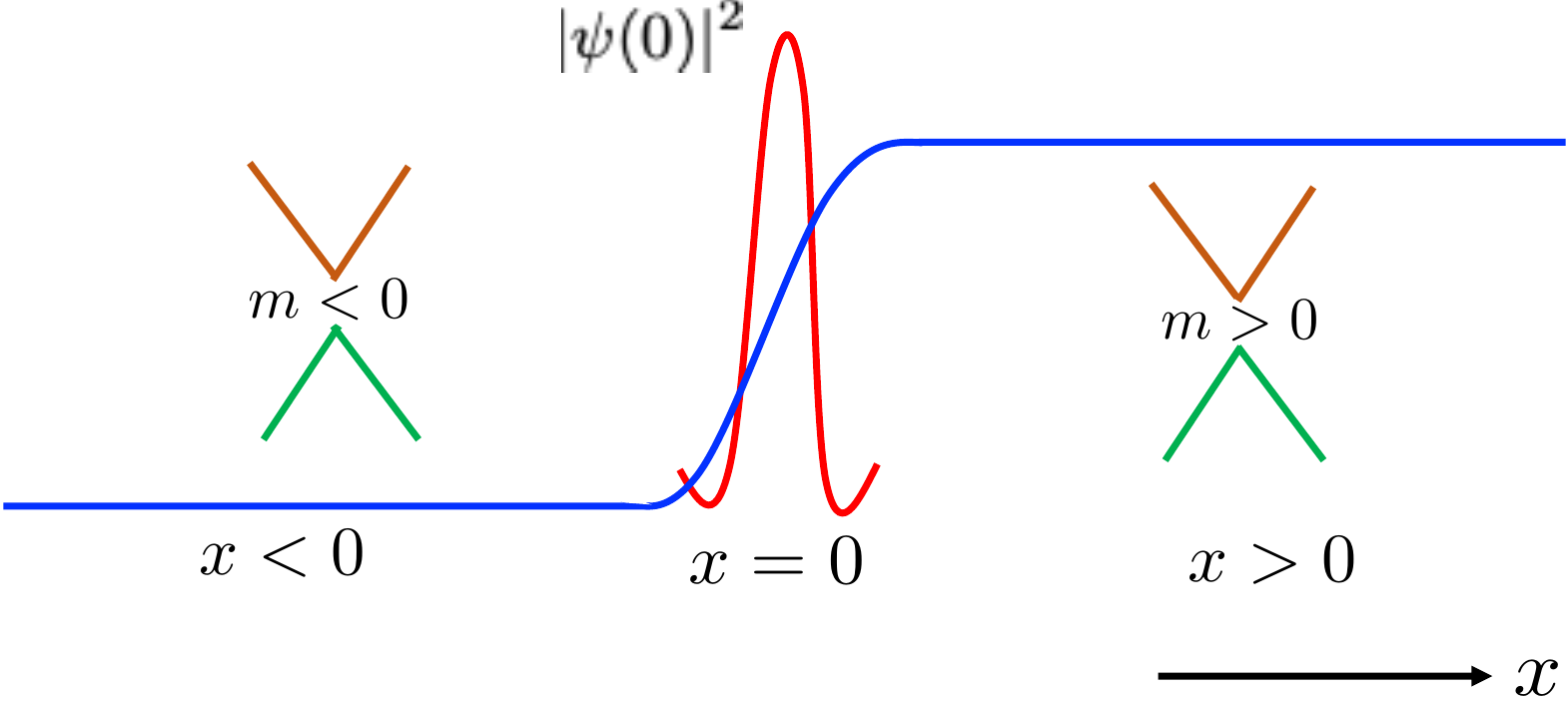}
\vskip-3.0cm
\caption{(Color online) In 1D, a Dirac Hamiltonian with spatially varying mass term hosts a zero-energy bound states (JR modes) localized at the boundary ($x=0$) where the mass term changes its sign.
One particular example of such mass term can be $m(x)=m\tanh(x/\xi)$ where, $\xi$ is the localization length of the mode.}
\label{Fig2}
\end{figure}
\vskip +2.5cm
At first, let us consider a particular example where the mass term has a step-function profile as $m(x)=m~{\rm sgn} (x)$ with $m>0$. This means $m(x <0) =-m$ and $m(x >0) = +m$. In this case,
there exists a solution at zero energy $E=0$, and the corresponding wave-function can be written as
\begin{eqnarray}
\Psi(x)=\sqrt{\frac{mv_{F}}{2\hbar}} 
\begin {pmatrix}
1\\
i
\end{pmatrix}e^{-\lvert m v_{F} x\rvert/\hbar}\ ,
\label{Eq2}
\end{eqnarray}
where $(1, i)^{T}$ is called the spinor. This solution is known as ``Jackiw-Rebbi'' mode and was first obtained by Jackiw and Rebbi in Ref.~\cite{JackiwRebbi}. The corresponding probability of this mode dominates near the interface at $x=0$ and decays exponentially away from that as shown in Fig.~\ref{Fig2}. The spatial distribution or localization length of JR zero-mode can be determined by the
characteristic length scale $\xi_{\pm}=\hbar/\lvert m v_{F}\rvert$ which indicates that this mode is sharply peaked when $m\rightarrow \infty$. 

While we present a particular example of position-dependent mass term for illustrative purpose, one can find a general solution of JR zero-energy mode for a distribution of mass $m(x)$ changing from
negative to positive at the two ends as illustrated in Fig.~\ref{Fig2}. One such example of mass term can be $m(x)=m\tanh(x/\xi)$. The general solution of the zero-energy mode can be written of the form
\begin{eqnarray}
\Phi(x) \propto e^{-\int_{0}^{x} m(x^{\prime})d x^{\prime}/\hbar v_{F}}\begin {pmatrix}
1\\
i
\end{pmatrix}\ .
\label{Eq3}
\end{eqnarray}
This indicates that the above zero-energy solution is robust and exists independent of the exact mass profile function and hence topological in nature. 
\subsection{2D HOTI and Corner Modes}
\begin{figure}[!thpb]
\centering
\includegraphics[width=0.55\linewidth]{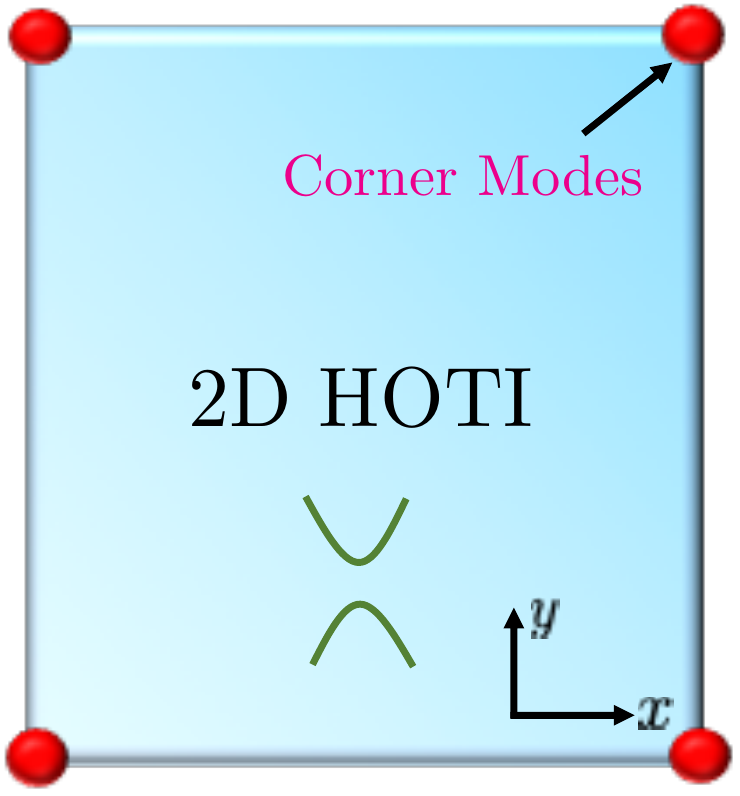}
\vskip-4.0cm
\caption{(Color online) Cartoon of a 2D HOTI (SOTI) in which both the bulk and edges are gapped and 0D corner modes (denoted by red bullets) appear at zero-energy located at the four corners 
of the system.
}
\label{Fig3}
\end{figure}

\begin{figure}[!thpb]
\centering
\includegraphics[width=1.0\linewidth]{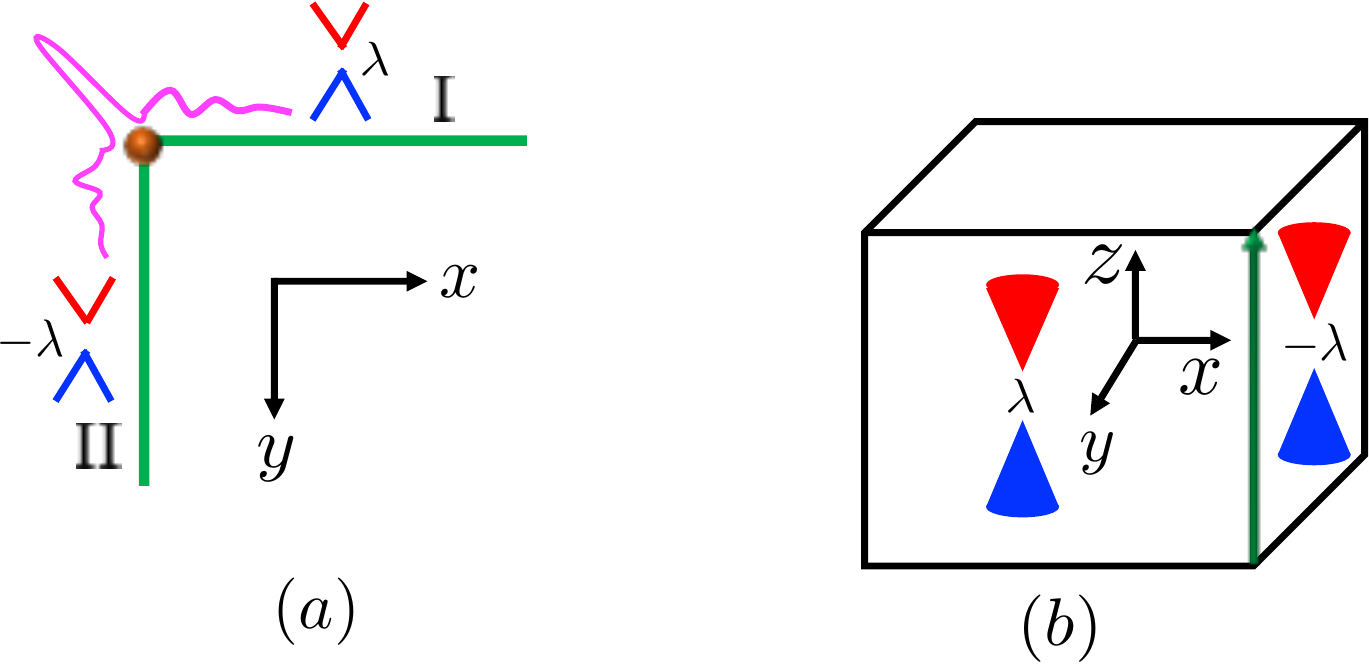}
\vskip-5.5cm
\caption{(Color online) (a) The Dirac mass terms ($\propto$ $\lambda$ and $-\lambda$), that change sign accross the corners leading to appearence of corner modes, are schematically shown. 
(b) The mass terms ($\propto$ $\lambda$ and $-\lambda$) are depicted for the Dirac surface electrons on the ($xz$ and $yz$) surfaces.}
\label{Fig4}
\end{figure}

A schematic diagram of 2D HOTI (SOTI in 2D) is demonstrated in Fig.~\ref{Fig3} in which both the bulk and edges are gapped and 0D topological zero-energy modes (denoted by red dots) appear at the 
corners of the sample. To realize this phase, we begin with a 2D TI, modeled in a square lattice with hopping elements $t_{x,y}$, spin-orbit coupling $\lambda_{x,y}$ ($x,y$ represent the two spatial directions), chemical potential $\mu$ and mass term $m_{0}$. Such mass term $m_{0}$ can appear in a material due to crystal-field splitting. This system preserves TRS and topological phase appear 
when $[m_{0}^{2} - (2t_{x}+2t_{y})^{2}][m_{0}^{2} - (2t_{x}-2t_{y})^{2}]<0$, hosting gapless propagating helical edge modes as shown in Fig.~\ref{Fig1}. 
\vskip +0.5cm
\vskip +3.5cm
In this system we now introduce a mass term of the form $\mathcal{H}_{m}=\lambda[\cos k_{x} - \cos k_{y}]$ where, $k_{x}$, $k_{y}$ are the wave-vectors along $x$, $y$ directions respectively and 
$\lambda$ is the strength of the mass perturbation. Here, ${\mathcal{H}}_{m}$ breaks both $\hat{{\mathcal{C}}}_{4}$ rotational symmetry and TRS $\hat{T}$. However, it preserves the combined 
symmetry operation $\hat{{\mathcal{C}}}_{4}\hat{T}$ which preserves the symmetry of the bulk. In presence of this mass perturbation, the 1D edge modes become gapped and one can find effective 
1D Dirac equation with mass term for the edges. For example, the Dirac masses on edges I and II carry opposite signs (proportional to $\lambda$ and $-\lambda$ respectively as shown 
in Fig.~\ref{Fig4}(a)), leading to
\vskip +0.5cm
\vskip +4.5cm
localized zero-energy modes (ensured by JR theory discussed before) at the intersection of two perpendicular edges (see Fig.~\ref{Fig4}(a)). The probability of the zero-energy modes 
is localized at the corners and decay exponentially along edges I (II) (see Fig.~\ref{Fig4}(a)) with spatial extent $\sim \hbar v_{F} \lambda_{x} /\lambda t_{y}$ ($\sim \hbar v_{F} \lambda_{y} /\lambda t_{x}$). 
The appearence of other corner modes can also be explained by the same mechanism. Although, the bulk of the system remains gapped. Thus, a 2D HOTI accomodating in-gap 0D corner modes 
at four corners can be realized in this square lattice system. 
\section{$\rm 3D$ Higher Order Topological Insulator {\label{sec:IV}}}
\begin{figure}[!thpb]
\centering
\includegraphics[width=0.9\linewidth]{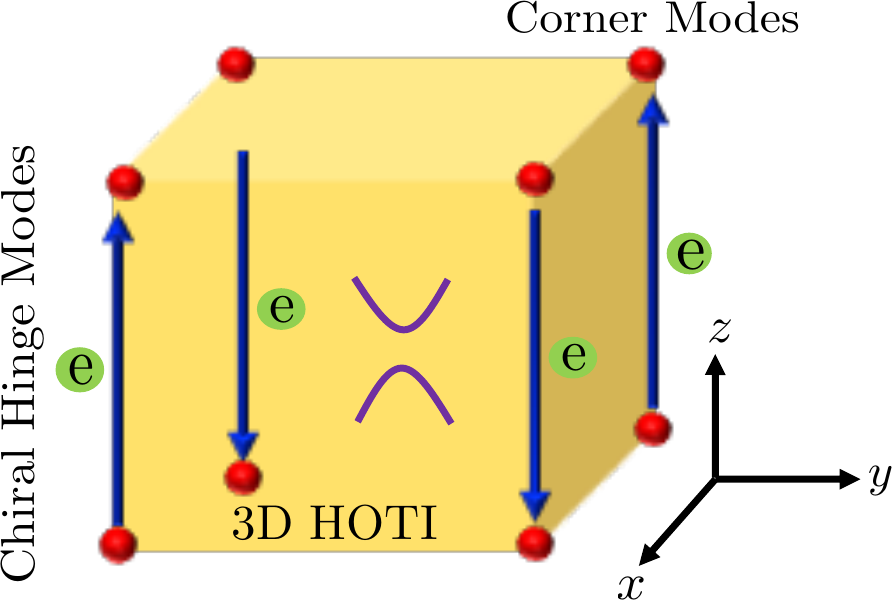}
\vskip-5.5cm
\caption{(Color online) Schematic of a 3D HOTI in which both bulk and surfaces are gapped and (a) 1D chiral propagating hinge modes (denoted by blue arrows) appear in the SOTI phase 
and (b) 0D corner localized modes (represented by red spheres) indicate the TOTI phase. 
}
\label{Fig5}
\end{figure}

Cartoon of a 3D HOTI is depicted in Fig.~\ref{Fig5} where both the bulk and surfaces are gapped hosting topological 1D propagating modes (SOTI in 3D) at the hinges or 0D localized modes (TOTI) 
at the corners of the sample. To realize these aspects, one can start with a cubic lattice of 3D TI. Similar to 2D TI, this lattice system also consists of hopping amplitude $t$, SOC strength $\lambda_{1}$
(for simplicity we assume $t$ and $\lambda_{1}$ to be isotropic  along the there spatial directions $x$, $y$, $z$) and mass term $M$. This system preserves 3D topological phase if $1<\lvert M/t \rvert <3$.
In presence of $\hat{{\mathcal{C}}}_{4}$ and $\hat{T}$ breaking mass term of strength $\lambda$ (see previous section for details), the system becomes a chiral 3D HOTI for $1<\lvert M/t \rvert <3$ 
and $\lambda_{1}, \lambda \neq 0$~\cite{Schindler}. Thus, this phase represents a SOTI in 3D hosting $z$-directed 1D propagating chiral hinge modes as illustrated in Fig.~\ref{Fig5}. 
\vskip -5.5cm
\hskip -6.8cm
\fbox{\begin{minipage}{15em}
This 3D HOTI is called ``chiral'' due to uni-directional (along $z$) propagating hinge modes present in this phase. In this case, the reason for obtaining chiral hinge state is the broken TRS in the bulk. 
\end{minipage}}
\vskip +2.0cm
\vskip +5cm
Physically, the main effect of mass perturbation $\lambda$ is to open band gaps with alternatiing signs for the surface Dirac electrons of the 3D TI on the $xz$ and $yz$ surfaces 
($\lambda$ and $-\lambda$ respectively as shown in Fig.~\ref{Fig4}(b)). The four hinge states are then domain walls at which the Dirac mass changes sign. It is evident from JR theory 
that such a domain wall on the surface of a 3D TI binds a gapless chiral mode, which, in this case can be interpreted as the hinge mode of the 3D SOTI~\cite{Schindler}. The wave-function
of the $z$-directed propagating hinge mode decays exponentially to the bulk along $x$ and $y$ directions. Note that, the hinge modes can also be of helical nature in some other context~\cite{Schindler}.

The mechanism for the realization of TOTI phase hosting 0D corner modes (see Fig.~\ref{Fig5}), is rather a complex subject and beyond the scope of the present general article. 

\section{Higher Order Topological Superconductor {\label{sec:V}}}
\begin{figure}[!thpb]
\centering
\includegraphics[width=1.2\linewidth]{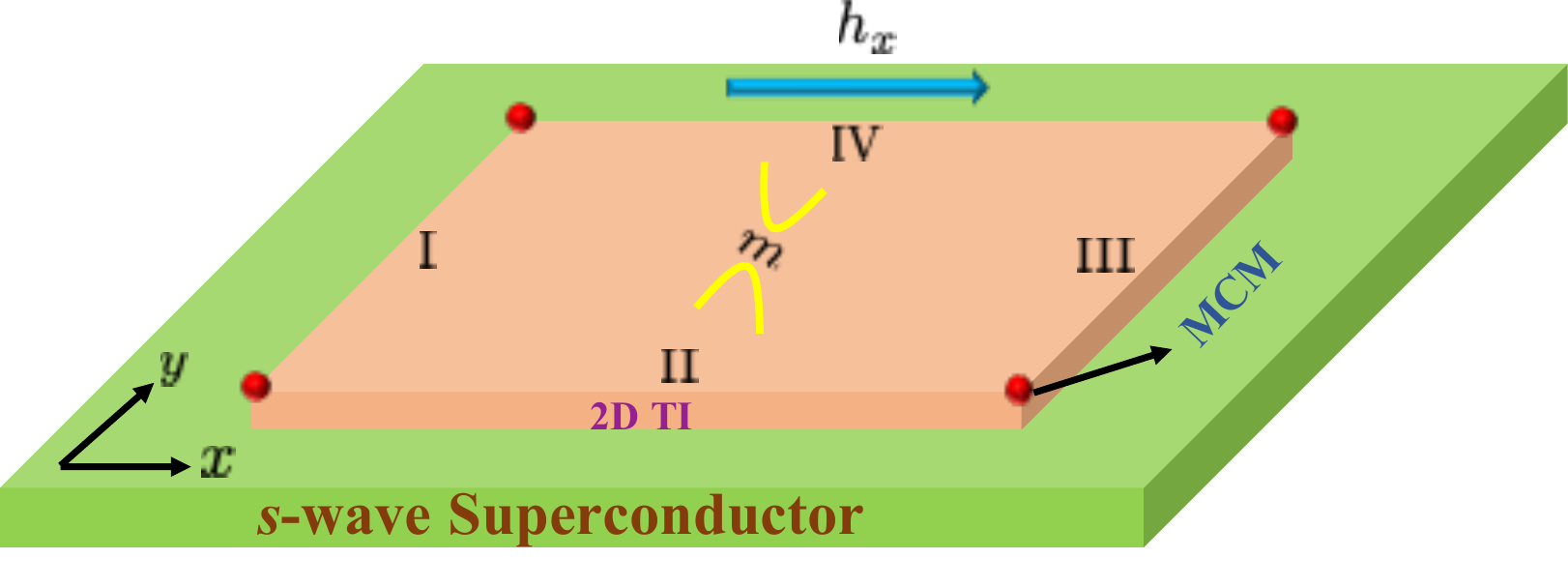}
\vskip -1.7cm
\caption{(Color online) Illustration of a heterostructure comprising of a 2D TI placed on top of an $s$-wave superconductor and subject to an in-plane Zeeman field. 
The four edges of the sample are denoted by I, II, III and IV. The red spheres at four corners of the system represent four localized MCMs.}
\label{Fig6}
\end{figure}

The bulk-boundary correspondence of HOTI can also be generalized for HOTSC hosting 0D MCMs and 1D MHMs. A schematic set-up for 2D HOTSC is demonstrated in Fig.~\ref{Fig6}
in which we start from a 2D TI in close proximity to a bulk $s$-wave superconductor. 
\vskip -6cm
\hskip +10.8cm
\fbox{\begin{minipage}{15em}
If a system is non-superconducting by itself, then superconductivity can be induced in it by placing it very close to or on top of a bulk superconductor. Here, the mechanism for inducing superconductivity
is tunneling of Cooper pairs from the bulk superconductor to the parent material. This process is known as ``proximity effect''. 
\end{minipage}}
\vskip +0.5cm
\vskip +1.0cm
Here, the parent 2D TI system is not a superconductor by itself. However, superconductivity can be induced in it
via the proximity effect. In presence of the proximity induced superconducting pairing gap $\Delta$, the edge (I, II, III, IV) 
spectrum of the sample becomes gapped and the system behaves as a trivial superconductor. The in-plane Zeeman field
$h_{x}$ exhibits different effects on the single particle edge spectra (\ie~$\Delta=0$) along $x$ and $y$ directions. 
In particular, $h_{x}$ can (cannot) open up the gap along edge-II (edge-I). Although, the Zeeman field breaks TRS, but the system 
is protected by the PH symmetry. Such anisotropic effect of $h_{x}$ leads to very different gapped spectrum 
when $\Delta \neq 0$. Along edge-II, the effective gap becomes $\Delta- h_{x}$ which means it can change sign at the
critical value $h_{x}=\Delta$. On the other hand, the Dirac mass along edge-I remains $\Delta$ and does not change sign. 
\begin{figure*}
\hskip -2.5cm
\centering
\includegraphics[width=1.25\linewidth]{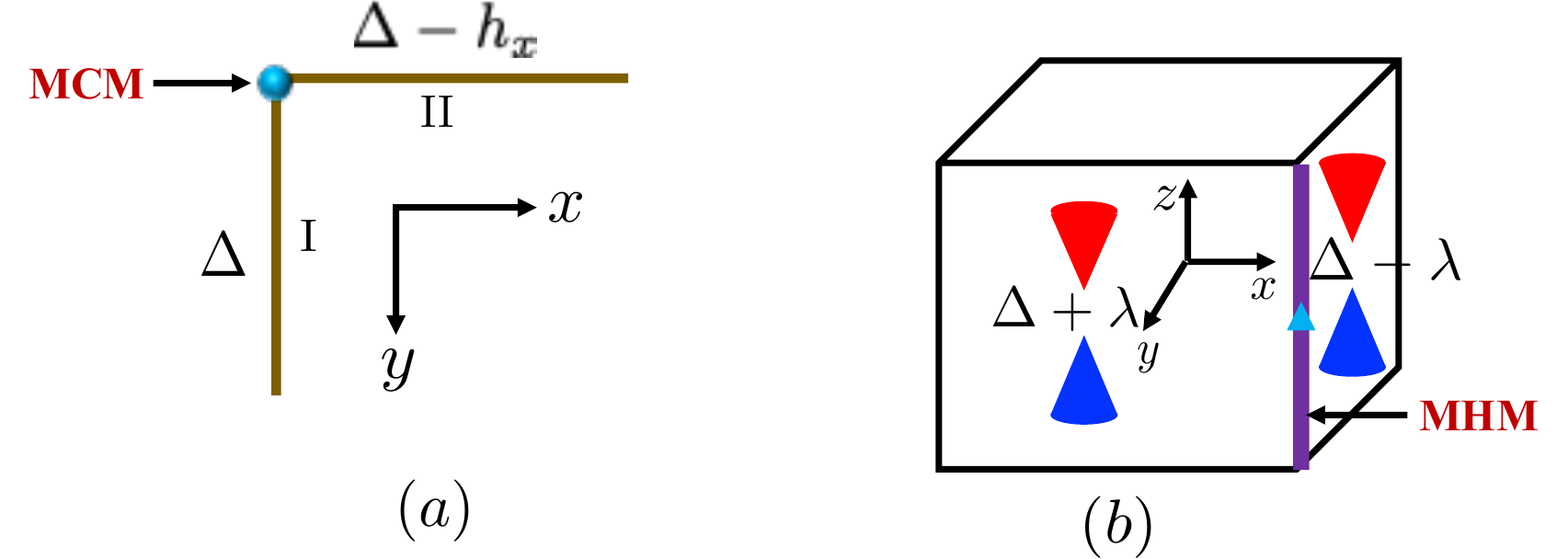}
\caption{(Color online) (a) The Dirac mass terms, $\Delta - h_{x}$ and $\Delta$ on edge-I and II respectively, change sign when $h_{x}>\Delta$ leading to the appearence of MCMs. 
(b) For $\lambda>\Delta$, mass terms of the surface Dirac electrons carry opposite signs which gives rise to chiral MHMs between the neighboring $xz$ and $yz$ surfaces.}
\label{Fig7}
\end{figure*}
This is schematically depicted in Fig.~\ref{Fig7}(a). Therefore, for $h_{x} > \Delta$, Dirac masses on edges-I and II carry opposite signs, 
leading to localized 0D MCMs at the intersection of two perpendicular edges (see Fig.~\ref{Fig7}(a)). In that sense, the emerging zero-energy 
MCMs can be interpreted as a special variant of the Jackiw-Rebbi zero mode protected by PH symmetry. The localization length of the MCM
can be different along $x$ and $y$ directions as $\hbar v_{F}/\lvert \Delta - h_{x} \rvert \neq \hbar v_{F}/\Delta$. Thus one can realize a second
order topological superconductor (SOTSC) hosting MCMs in this 2D system. 

\begin{figure*}
\hskip -1.0cm
\centering
\includegraphics[width=1.2\linewidth]{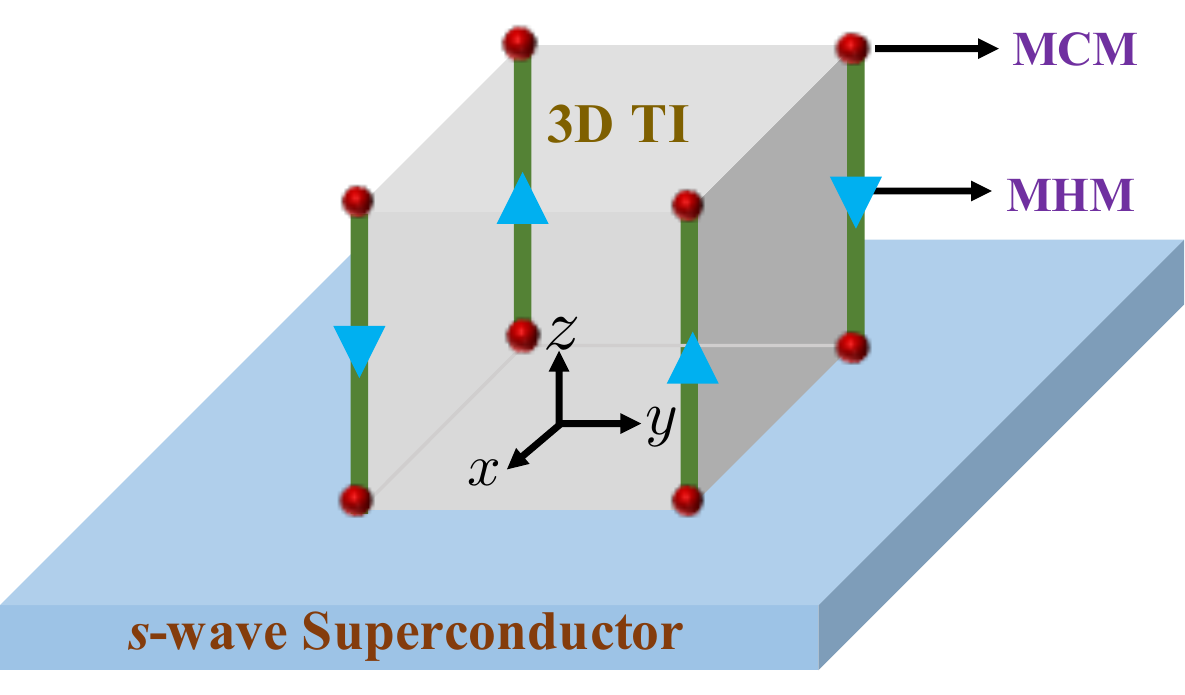}
\caption{(Color online) Demonstration of a schematic set-up for the realization of 3D HOTSC in which a cubic 3D TI is placed in close proximity to a bulk $s$-wave superconductor. 
Chiral MHMs are present in the SOTSC phase, as marked by the green lines. MCMs appear in the TOTSC phase, as depicted by the red spheres.}
\label{Fig8}
\end{figure*}

In a similar fashion, 3D HOTSC can be realized in a set-up depicted in Fig.~\ref{Fig8}, where a 3D TI is placed in close proximity to a $s$-wave
superconductor. In presence of the superconducting pairing $\Delta$, the boundary 2D surface states become gapped. Incorporating the 
$\hat{{\mathcal{C}}}_{4}$ rotational symmetry and TRS $\hat{T}$ broken mass term $\mathcal{H}_{m}=\lambda[\cos k_{x} - \cos k_{y}]$, 
the mass gap for the surface Dirac electrons can change sign between the two neighboring surfaces as depicted in Fig.~\ref{Fig7}(b). Here, 
for $\lambda>\Delta$, the mass gap on the $xz$ and $yz$ surfaces carry opposite sign and this results in 3D SOTSC hosting 1D chiral MHMs 
propagating along the $z$ direction. The same can also be achieved by applying a Zeeman field (not necessarily in-plane) in the sample. 

Similar to TOTI, the discussion on emergence of third order topological superconductor (TOTSC) phase (see Fig.~\ref{Fig8}), accommodating 0D MCMs, 
is beyond the scope of the present article. 

\section{Realization of Higher Order Topological Systems {\label{sec:VI}}}
As far as experimental progress of higher-order topological (HOT) systems are concerned, these phases was first experimentally realized in a classical gigahertz-frequency reconfigurable microwave 
circuit~\cite{Peterson}, electrical circuit~\cite{Imhof} and in a mechanical metamaterial (a material with tunable mechanical properties)~\cite{Garcia}. Then 2D SOTI is experimentally realized in acoustic system, based on a ``breathing Kagome lattice''~\cite{HXue} and dimensional hierarchy of higher-order topology (1D hinge states and 0D corner modes) is experimentally discovered in single 3D simple-cubic acoustic sonic crystals~\cite{XZhang}. However, the search for the HOT phases in solid state materials is still in its infancy. In recent times, F. Schindler \etal~in Ref.~\cite{FSchindler} have established that Bismuth (Bi) is in fact a HOTI. Their claim is supported by both theoretical analysis and complimentary experimental techniques. Very recently, $\rm Td-WTe_{2}$ is experimentally predicted to host 
HOT phases with topologically protected, helical 1D hinge states~\cite{YBChoi}. Moreover, experimental evidence of HOTI, in a real 3D material, has been found in Bismuth-Bromide ($\rm Bi_{4}Br_{4}$)
via ARPES measurements~\cite{Noguchi}. However, note that the set-ups that we discuss in the context of HOTI and HOTSC in this present article, are still to be realized from the experimental point of 
view in real materials. 

\section{Conclusions and Outlook {\label{sec:VII}}}
In this general article, we have provided a pedagogical introduction to the new emerging field of HOTI and HOTSC in quantum condensed matter physics. In these intriguing HOT phases, 
gapless boundary modes dwell on $(d - n)$-dimensional boundaries of a $d$-dimensional system, unlike $(d - 1)$-dimensional boundary of a first order TIs. We discuss different set-ups
and possible symmetry breaking perturbations which can give rise to 2D and 3D SOTI hosting zero-energy corner localized modes and propagating hinge modes respectively. Furthermore, 
we emphasize various set-ups that in presence of proximity induced superconductivity and magnetic field, can host HOTSC (we mainly discuss SOTSC in 2D and 3D) phase anchoring 0D MCMs 
and 1D MHMs. Finally, we briefly present the experimental development in this field based on classical systems and solid-state material perspective. However, we have not discussed some
challenging issues such as the classification of HOT systems~\cite{Geier,Luka} and determination of the appropriate topological invariants that distinguish them from conventional first order TIs~\cite{Benalcazar1,Benalcazar2,Resta}; we direct the reader to the original articles for details. 

In HOT systems, one of the prime interests is to generate such phases via external periodic driving (for \eg~laser) starting from a trivial (non-topological) system. Another interesting direction
is to generalize these ideas from the current setting of non-interacting fermionic systems to strongly correlated fermions or bosons. The effects of strong disorder and possible realization of
HOT Anderson insulator or superconductor is also a prime area of interest. On the experimental side, fabricating different set-ups of fermionic systems to realize HOT phases still remains a 
challenging task. Also, distinguishing hinge modes from the edge states via transport signal is woth exploring. From the application point of view, 0D MCMs in 2D and 3D HOT systems can be 
more beneficial in fault-tolerant topological quantum computation, compared to their 1D NW MZMs counterpart, as far as tunability and braiding statistics are concerned. Also, the topological 
propagating hinge modes can be potential candidate towards future spintronics applications. All in all, there are still surprises in store as we probe deeper into the realm of topological quantum matter. 

\section{Acknowledgement}

One of us (AMJ) thanks DST, India for financial support (through J. C. Bose National Fellowship).

\end{document}